\documentclass[letterpaper,floatfix,aps,pra,amsmath,amssymb,twocolumn,superscriptaddress,
showpacs]{revtex4-1}

\usepackage{epsfig}
\usepackage{verbatim}
\usepackage{color}
\usepackage{epsf}
\usepackage{array}
\usepackage[pdfauthor={Riwar, Glazman, Catelani},
            pdftitle={Dissipation by normal-metal traps in transmon qubits}]{hyperref}

\hypersetup{pdfstartview={XYZ null null 1.05}}

\newcommand{\be}{\begin{equation}}
\newcommand{\ee}{\end{equation}}
\newcommand{\bea}{\begin{eqnarray}}
\newcommand{\eea}{\end{eqnarray}}

\renewcommand{\Re}{\mathrm{Re}\,}

\newcommand{\eref}[1]{Eq.~(\ref{#1})}
\newcommand{\esref}[1]{Eqs.~(\ref{#1})}
\newcommand{\rref}[1]{(\ref{#1})}

\newcommand{\ocite}[1]{Ref.~\onlinecite{#1}}

\newcommand{\qp}{\mathrm{qp}}

\usepackage{epstopdf}
\begin{document}

\title{Dissipation by normal-metal traps in transmon qubits}

\author{R.-P. Riwar}

\affiliation{JARA Institute for Quantum Information (PGI-11), Forschungszentrum J\"ulich, 52425 J\"ulich, Germany}
\affiliation{Departments of Physics and Applied Physics, Yale University, New Haven, CT 06520, USA}

\author{L. I. Glazman}

\affiliation{Departments of Physics and Applied Physics, Yale University, New Haven, CT 06520, USA}

\author{G. Catelani}

\affiliation{JARA Institute for Quantum Information (PGI-11), Forschungszentrum J\"ulich, 52425 J\"ulich, Germany}

\begin{abstract}
Quasiparticles are an intrinsic source of relaxation and decoherence for superconducting qubits. Recent works have shown that normal-metal traps may be used to evacuate quasiparticles, and potentially improve the qubit life time. Here, we investigate how far the normal metals themselves may introduce qubit relaxation. We identify the ohmic losses inside the normal metal and the tunnelling current through the normal metal-superconductor interface as the relevant relaxation mechanisms. We show that the ohmic loss contribution depends strongly on the device and trap geometry, as a result of the inhomogeneous electric fields in the qubit. The correction of the quality factor due to the tunnelling current on the other hand is highly sensitive to the nonequilibrium distribution function of the quasiparticles. Overall, we show that even when choosing less than optimal parameters, the presence of normal-metal traps does not affect the quality factor of state-of-the-art qubits.
\end{abstract}

\date{\today}


\maketitle

\section{Introduction}

Superconducting qubits based on Josephson junctions are among the prime candidates to realize large scale quantum computing~\cite{Devoret_2004,koch2007,Manucharyan_2009}. Theory~\cite{lutchyn,prl,leppa} and experiment~\cite{shaw,martinis,3dtr,riste} agree that quasiparticles, the intrinsic excitations in a superconductor, provide a natural bound on the coherence lifetimes of such qubits. Tunnelling couples quasiparticles to the junction's phase difference, and thus introduces a relaxation process with a rate proportional to the quasiparticle density~\cite{prb1}.
Importantly, there is convincing experimental evidence that at low temperatures, this density is much higher than the expected thermal equilibrium value, indicating a residual, nonequilibrium quasiparticle population~\cite{martinis,riste}.
This is why it is paramount to provide means to evacuate the excess quasiparticles from the active region of the device. The so far studied strategies involve gap engineering~\cite{cpt,sun}, introducing vortices~\cite{ullom,plourde,wang,pekola2}, normal-metal traps~\cite{court,raja1,raja2,Riwar,Patel_2016,trapopt}, and recently pumping via control pulses~\cite{Gustavsson1573}. The first three approaches all share a common idea: to provide a subgap density of states into which quasiparticles can relax. Here we focus on normal-metal traps.

Trapping quasiparticles through the addition of small normal-metal layers on top of the superconducting one was demonstrated to be effective~\cite{Riwar}. A critical trap size was identified, above which the decay rate of quasiparticles is limited by the quasiparticle diffusion. In this diffusion-limited regime, the geometry of device and trap are important, as was also further elaborated in Ref.~\onlinecite{trapopt}.
The escape process whereby quasiparticles leak from the normal metal trap back into the superconductor was properly taken into account. As a result, the decay rate can be effectively reduced, and the transport at the N-S junction gives rise to a nonequilibrium distribution of excitations in the normal metal.

While normal metal traps provide a promising remedy against nonequilibrium quasiparticles, the traps \textit{themselves} may give rise to dissipation. For instance, subgap states induced near a junction by the inverse proximity effect can lead to a possible reduction of the qubit $T_1$ relaxation time~\cite{proximity}. This effect can however be easily made small by placing the trap further than several coherence lengths away from the active elements of the qubit. In contrast, in the present work we investigate qubit relaxation caused by processes taking place at the trap location.  We identify two main contributions. On one hand, a photo-assisted tunnelling current through the N-S interface gives rise to energy relaxation. On the other hand, currents inside the normal metal, due to the redistribution of charges in the ac electric field, give rise to ohmic losses.

The former depends on the nonequilibrium distribution function of quasiparticles -- be it in the superconductor or in the trap. It leads to a break-down of the fluctuation dissipation theorem, such that the quality factor has to be computed directly through the qubit lifetime, and cannot be inferred from the dissipated power.
As for the ohmic losses inside the normal metal, we find that due to the inhomogeneous electric field of the transmon, the trap geometry and placement may have a strong impact on the dissipated power. In some parameter regimes, a simple circuit picture with lumped elements fails to accurately describe the relaxation processes due to the normal-metal trap dissipation.
All in all, we show however, that even for a poor design choice, normal-metal traps do not appreciably affect the quality factor of the best transmons currently available.

The paper is organized as follows: in Sec.~\ref{sec:circuit_pic} we introduce a lumped element model for the qubit with a trap which includes dissipative elements. In Sec.~\ref{sec:normal_currents} we study ohmic dissipation by currents inside the normal metal and estimate the impact of the dissipation on the transmon lifetime. In Sec.~\ref{sec:decoherence} we calculate the rates of qubit transitions due to tunnelling between superconductor and normal metal using Fermi's golden rule. We argue that the effect on qubit lifetime is negligible compared to that of quasiparticle tunnelling through the qubit Josephson junction, see \eref{Gdqpr}. We summarize our work in Sec.~\ref{sec:summary}. Some details and extensions are presented in Appendices~\ref{app:conformal} through \ref{app:ns_diss}.

\section{Circuit model}
\label{sec:circuit_pic}

\begin{figure}
\centering
\includegraphics[width=0.3\textwidth]{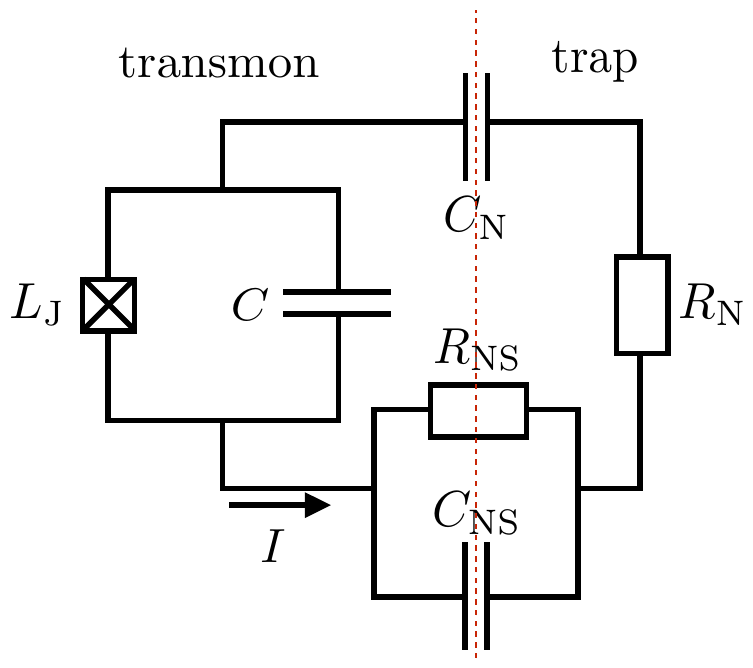}
\caption{Circuit scheme for the normal metal trap. The N-S tunnel junction
is modeled as a resistance $R_{\text{NS}}$ and a capacitance $C_{\text{NS}}$
in parallel. The metal itself has the resistance $R_{\text{N}}$ and
is capacitively coupled to the circuit with $C_{\text{N}}$. \label{fig:circuit-scheme}}
\end{figure}

We begin by considering a circuit picture for the qubit-trap system, taking for concreteness a single-junction transmon as the qubit, see Fig.~\ref{fig:circuit-scheme}.
The transmon consists of a junction with Josephson inductance $L_\text{J}$ and a capacitance $C$ between the two superconducting plates.
Since the transmon is only weakly anharmonic, in this section we treat it as an $LC$ resonator
with resonant frequency $\omega_0=1/\sqrt{L_\text{J}C}$.
To describe the trap, we introduce
a capacitance $C_\text{NS}$ for the N-S junction
and a capacitance $C_\text{N}$ between the normal metal and the other S plate.
To account for ohmic losses inside the normal metal we add a resistive element $R_\text{N}$ in series between the two capacitors, while the resistance $R_\text{NS}$
in parallel to $C_\text{NS}$ accounts for the losses due to the tunnel current.

The ideal lossless circuit is obtained by setting $R_\text{N} =0$ and $R_\text{NS} \to \infty$. In this limit, there are two superconducting island and a normal one, so the Lagrangian ${\cal L}$ for the circuit has two degrees of freedom, the phase differences $\varphi$ across the junction and $\phi$ across $C_\text{NS}$. In terms of these variables we have
\be\label{eq:lagr}
{\cal L} = \frac12 C \frac{\dot\varphi^2}{(2e)^2} + \frac12 C_\text{NS} \frac{\dot\phi^2}{(2e)^2} +\frac12 C_\text{N} \frac{\left(\dot\varphi - \dot\phi\right)^2}{(2e)^2} + E_\text{J} \cos\varphi
\ee
where $E_\text{J}=\left(\Phi_0/2\pi\right)^2/L_\text{J}$ is the Josephson energy, with $\Phi_0=h/2e$ the flux quantum. Since $R_\text{NS} \to \infty$ corresponds to neglecting N-S tunnelling, charge on the N island is conserved and we can eliminate the variable $\phi$, arriving at the usual transmon Lagrangian ($C_t \dot\varphi^2/8e^2 + E_\text{J} \cos\varphi$) with a total capacitance $C_t= C+C_\text{N}C_\text{NS}/(C_\text{N}+C_\text{NS})$. In other words, neglecting losses the trap only leads to a renormalization of the charging energy.

Keeping the resistive elements finite to phenomenologically account for losses, the resonant frequency of the coupled transmon-trap device can be computed as the zero of the total circuit admittance, leading to the equation
\be
\omega^{2}-\omega_{0}^{2}=\frac{i\omega}{CZ_{\text{trap}}\left(\omega\right)}\ .
\ee
where the total impedance of the trap is
\be
Z_{\text{trap}}(\omega)=\frac{1}{i\omega C_{\text{N}}}+R_{\text{N}}+\frac{1}{\frac{1}{R_{\text{NS}}}+i\omega C_{\text{NS}}}\ .
\ee
For a trap impedance larger than the transmon one, $Z_\text{trap}(\omega_0) \gg Z_q \equiv \sqrt{L_\text{J}/C}$, we find that the resonant frequency is shifted by a small (complex) amount
\be
\delta\omega=\frac{i}{2CZ_\text{trap}(\omega_0)}\, .
\ee
This shift determines the quality factor as
\be
Q=\frac{\omega_0}{2\text{Im}[\delta\omega]} = \left[\Re \frac{Z_q}{Z_\text{trap} (\omega_0)}\right]^{-1} \, .
\ee
The quality factor, due to the assumption $Z_\text{trap} \gg Z_q$, is much larger than one. In this model, $Q$ would diverge for the ideal qubit, $R_\text{N} \to 0$ and $R_\text{NS} \to \infty$. Taking into account small but finite dissipation, we make the simplifying assumptions
(to be discussed in what follows) $\omega_0 R_\text{N} C_\text{N} \ll 1$, $\omega_0 R_\text{NS} C_\text{NS} \gg 1$, and $C_\text{N} \ll C_\text{NS}$ to find
\be
Q^{-1} \simeq \frac{C_\text{N}}{C} \left( \omega_0 R_\text{N} C_\text{N} + \frac{C_\text{N}}{C_\text{NS}} \frac{1}{\omega_0 R_\text{NS} C_\text{NS}}\right) \, .
\ee
This expression has the form of the sum of the inverse quality factors of the series combination of $R_\text{N}$ and $C_\text{N}$ plus the parallel combination of $R_\text{NS}$
and $C_\text{NS}$, with the two terms weighted by appropriate participation ratios determined by the capacitors. The same formula can be obtained by calculating the ratio between the sum of the two dissipated powers per cycle in the two resistive elements, $P_\text{N}$ plus $P_\text{NS}$ over $\omega_0$, and energy $E$ stored in the circuit:
\be\label{QP}
Q^{-1} = \frac{P_\text{N} + P_\text{NS}}{\omega_0 E}\, .
\ee

In terms of the relaxation time $T_1$, the qubit quality factor $Q_q$ is defined as
\be\label{Qq}
Q_q = T_1 \omega_{0}
\ee
The equivalence between the ``circuit'' quality factor of \eref{QP} and $Q_q$ rest on the use of fluctuation-dissipation relations, valid in thermal equilibrium.
For example, the effect of a shunting resistor (such as $R_\text{N}$) on the qubit relaxation rate has been evaluated before using
the Caldeira-Leggett, spin-boson model -- see \textit{e.g.} the review \cite{rmp73} -- in which the resistor is described as an equilibrium bosonic bath. The result agrees with the calculation presented above, meaning that we can indeed just calculate the power dissipated by the normal metal, as we will do in the next section, to arrive at its contribution to the qubit relaxation rate. For the tunnelling between S and N, on the other hand, we cannot in general assume thermal equilibrium, since there is firm evidence that at low temperatures quasiparticles in a superconductor are not in equilibrium~\cite{riste,klapwijk}. Therefore in Sec.~\ref{sec:decoherence} we will directly calculate the contribution due to N-S tunnelling to the qubit relaxation time using Fermi golden rule.

\section{Dissipation by currents inside the normal metal}
\label{sec:normal_currents}

\begin{figure}[t]
\includegraphics[width=0.42\textwidth]{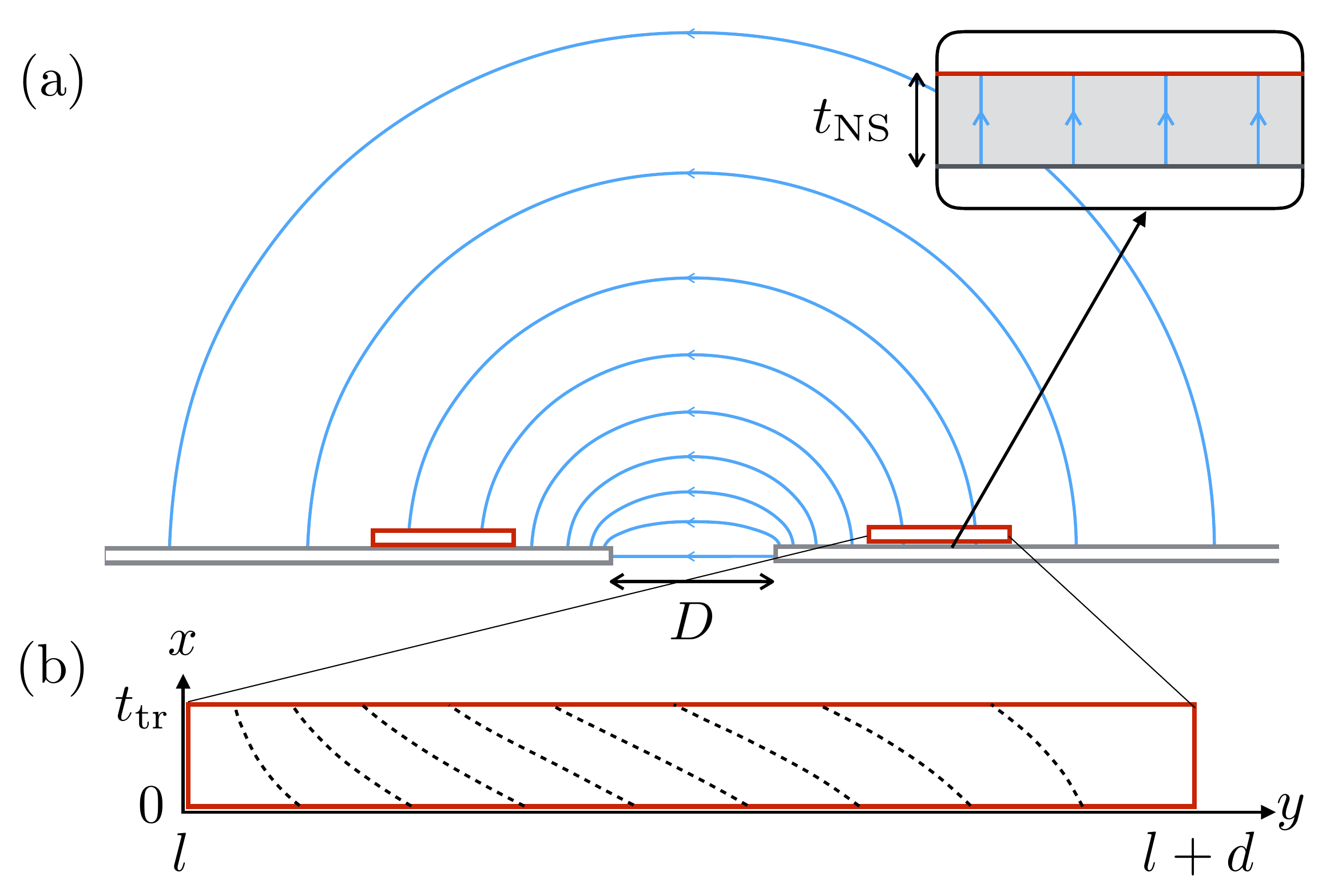}\caption{(Color online) (a) Sketch of the field lines (blue, with arrows) of the transmon with trap. The transmon
itself behaves as a coplanar capacitor, with strong fields at the
edges close to the Josephson junction, and a decaying field far away.
The transmon plates are separated by $D$, the identical traps are placed a distance $l$
away from the edges, and have length $d$. Inset: the N-S junction can be
considered as a parallel plate capacitor with a homogeneous field inside the insulating layer (gray)
of thickness $t_{\text{NS}}$. (b) Zoom into the normal metal. The dashed lines correspond to field lines of the current density $\vec{j}(x,y)$, thus indicating the path of probe charges from top to bottom (and vice versa) in presence of an ac voltage. When moving upwards, charges shift to the left. Therefore, in addition to the current in $x$-direction, there is a displacement of charges along the $y$-direction.}\label{fig:transmon_capacitor}
\end{figure}

In the circuit picture of the preceding section we have simply assigned a resistance $R_\text{N}$ to the normal-metal trap. However, it is not immediately clear how to relate
this phenomenological resistance to material properties (such as the resistivity $\rho_\text{N}$) and the device geometry. If the qubit capacitor were in the simple parallel plate configuration, with the trap covering a small part of one plate, the electric field lines would be homogenous (neglecting fringe fields) and the charges in the normal metal would move across the thickness of the trap. Therefore, the relevant resistance entering in $P_\text{N}$ would be proportional to $\rho_\text{N} t_\text{tr}$, with $t_\text{tr}$ the normal-metal trap thickness. A realistic qubit design, however, has a coplanar geometry, and the electric field is non-homogenous, see Fig.~\ref{fig:transmon_capacitor}a. As a consequence, charges move both across the thickness ($x$ direction in Fig.~\ref{fig:transmon_capacitor}b) and parallel to the surface ($y$ direction), with the latter contributing to $P_\text{N}$ a term proportional to $\rho_\text{N}/t_\text{tr}$ which can be dominant in thin films.

\begin{figure}[b]
\includegraphics[width=0.46\textwidth]{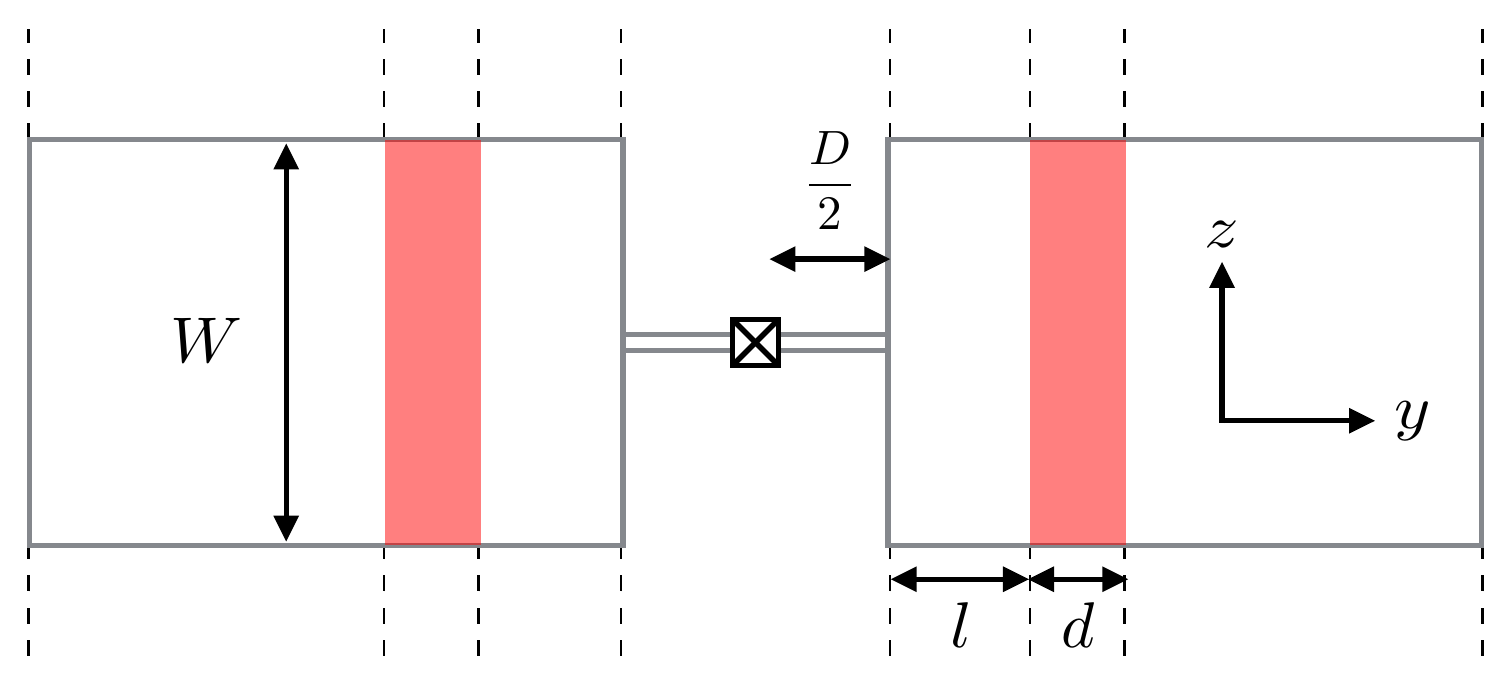}
\caption{(Color online) Typical geometry for 3D transmons~\cite{3dtr} with large pads of width $W$ at a distance $D/2$ from the junction. Traps (light red) have size $d$ and are placed at a distance $l$ from the pads edges. Dashed lines denote extension to infinity to enable analytical calculations -- see text for details.}\label{fig:transmons}
\end{figure}

A typical 3D transmon geometry is depicted in Fig.~\ref{fig:transmons}, where two coplanar capacitor plates of width $W$ are separated by distance $D$ in the $y$ direction. To estimate the role of traps on qubit coherence, we consider two normal-metal strips of widths $d$ placed symmetrically at a distance $l$ from the edge of the qubit capacitor plates; to make analytical calculation possible, we treat the capacitor plates and normal metal as extending to infinity, as indicated by the dashed lines in Fig.~\ref{fig:transmons}.
Then the coplanar capacitor consists of two semi-infinite plates, placed in the $y$-$z$ plane;
a cross-section is shown in Fig.~\ref{fig:transmon_capacitor}. Neglecting for now the normal-metal islands, the surface charge density on a capacitor plate is (see Appendix~\ref{app:2Dplates})
\be\label{sigma_y}
\sigma\left(y\right)=\frac{\epsilon_{0}V}{\pi}\frac{1}{\sqrt{y\left(y+D\right)}}
\ee
where $V$ is the voltage difference between the plates, $\epsilon_0$ is the vacuum permittivity, and $y=0$
is position of the edge of the right plate (without loss of generality
we choose here to consider the right plate in Fig.~\ref{fig:transmon_capacitor}). Close
to the edge the charge density diverges as $y^{-1/2}$, whereas
far away, $y\gg D$, the surface charge decays as $y^{-1}$. This sufficiently fast decay justifies the extension of the capacitor plates to infinity in the $y$ direction. Similarly, we neglect the finite size of the transmon in the $z$ direction, as it will not bring substantial modifications so long as the plates are much wider than their distance, $W \gg D$.
Finally, we note that the voltage $V$ appearing in \eref{sigma_y} depends on time as
$V\left(t\right)=V_{0}\cos\left(\omega_0 t\right)$. In writing these expressions we assume that the transmon frequency $\omega_0$ is sufficiently
low, so that the electric currents adjust to the new voltage instantaneously; this assumption is justified since typical qubit frequencies ($\lesssim 10\,$GHz)
are much smaller than the plasma frequency in a metal ($\sim 10^6\,$GHz).

Let us now include traps, assuming that their thickness $t_\text{tr}$ is negligible with respect to $D$. Furthermore, the condition $\omega_0 R_\text{NS}C_\text{NS} \gg 1$ already mentioned in Sec.~\ref{sec:circuit_pic} means that the capacitor formed by the trap and the superconductor does not appreciably discharge during an oscillation period.
Thus, we consider as a first approximation the limiting case where no charges are exchanged between transmon and trap. In this case, Eq.~\eqref{sigma_y} represents the surface charge on the top surface of the transmon for all $y$ where there is no trap and on the top surface of the normal-metal. The interface between metal and superconductor can be treated as a parallel plate capacitor with a homogeneous field (neglecting
fringe effects), see inset of Fig.~\ref{fig:transmon_capacitor}a. This is justified since the thickness of the oxide layer separating N and S is of the order of 1~nm, much less than the trap dimensions in the other directions (of the order of $10~\mu$m or more). With charge conservation in the normal metal, the (homogeneous)
surface charge density on the bottom surface of the trap is
\be\label{eq:sigmap}
\sigma'=-\int_{l}^{d+l}\frac{dy}{d}\sigma\left(y\right).
\ee

Using the above assumptions, we can compute the current density $\vec{j}(x,y)$ in the normal metal. As we show in Appendix~\ref{app:Ncurr}, the two components $j_x$ and $j_y$ can be expressed in terms of the function $\sigma_y$ of \eref{sigma_y} and its integral. The dissipated power can be written in the form $P_\text{N} = P_x + P_y$, where the contributions from the current components are
\be\label{eq_P_i}
P_{i}=W\rho_\text{N}\int_{0}^{t_{\text{tr}}}\!dx\int_{l}^{l+d}\!dy\,j_{i}^{2}\left(x,y\right) \,.
\ee
Here $i=x$ or $y$; $\rho_\text{N}$ and $W$ are, respectively,  the resistivity (assumed isotropic) of the trap and its width in the $z$ direction.
We discuss in the following some of the relevant regimes of trap size and position.
In order to efficiently evacuate quasiparticles, and to reduce the density of quasiparticles at the junction, it is favourable to place sufficiently large traps close to the junction, see \ocite{trapopt}. Taking the limit of large traps, $d\gg D, \, l$, corresponds to the worst-case scenario in terms of the dissipated power -- this will provide an upper bound for the dissipation by currents inside normal metal. In this regime we find
\be\label{eq:Px_wcs}
P_x \simeq P_0 \frac{t_\text{tr}}{3D} \ln\left(1+\frac{D}{l}\right)
\ee
and
\be\label{eq:Py_wcs}
P_y  \simeq P_0 \frac{d}{t_\text{tr}} \mathrm{f}\left(\frac{d}{\max\{l,D/4\}}\right) ,
\ee
where
\be\label{eq:P0}
P_0 = \left(\frac{\epsilon_0 \omega_0 V_0}{\pi}\right)^2 \rho_\text{N} W
\ee
and the function $\mathrm{f}$ is
\be
\mathrm{f}(x)  = \frac13 \ln^2(x) -\frac32 \ln(x)+2 \, .
\ee
Note that $P_x$ in \eref{eq:Px_wcs} diverges in the limit $l\to 0$. This divergence originates from the square root singularity in \eref{sigma_y}; it is regularized by the finite thickness $t_\text{S}$ of the superconducting plate via the substitution $l \to t_\text{S}/4\pi$, see Appendix~\ref{app:regular}. Then for typical parameters (e.g., $d \sim 200\,\mu$m, $D\sim 50\, \mu$m, $t_\text{S} \sim 30\,$nm) we find that the logarithmic factors can be dropped for an order-of-magnitude estimate, so that for a trap at the edge of the capacitor we get
\be\label{eq:pxy_wsc}
P_x \approx P_0 \frac{t_\text{tr}}{D} \, , \quad P_y \approx P_0 \frac{d}{t_\text{tr}} \, .
\ee
Since $P_x/P_y \approx t_\text{tr}^2/dD$ is much smaller than unity, the loss is, remarkably, dominated by the parallel component $P_y$.

For a small trap close to the edge, $l \ll d \lesssim D$, we find
\bea
P_x & \simeq & P_0 \frac{t_\text{tr}}{3D} \left[\ln \frac{d}{l} + 8 \right] \, , \\
P_y & \simeq & P_0 \frac{d}{t_\text{tr}} \frac{2}{15}\frac{d}{D}. \label{eq:py_se}
\eea
For $P_x$ the regularization discussed above still applies, and for typical parameters we estimate the ratio $P_x/P_y \lesssim 30 (t_\text{tr}/d)^2$ to be again much smaller than unity (here it should be kept in mind that in practice a trap is at least of $\mu$m length).

Finally, let us consider a small trap far from the edge, $d, \, D \ll l$, which minimizes the dissipated power. In this case we find
\bea\label{eq:px_f}
P_x & \simeq & P_0 \frac{t_\text{tr} d}{l^2} \, ,\\
P_y & \simeq & P_0 \frac{d}{t_\text{tr}} \frac{1}{120}\left(\frac{d}{l}\right)^4. \label{eq:py_f}
\eea
Here, in contrast to the previous cases, the dissipated power can be dominated by the normal component $P_x$ when $d \ll \sqrt{10 t_\text{tr} l}$. As we show in Appendix~\ref{app:lest}, in this regime a lumped-element approach suffices to calculate the dissipated power.

In a recent article~\cite{trapopt}, we studied the effect of multiple traps with regard to quasiparticle evacuation. Splitting one large trap into multiple smaller ones (while conserving the total trap area) and distributing them evenly over the device is highly advantageous for the evacuation efficiency. In the same spirit, we here briefly consider the effect of a trap splitting on the losses.

\begin{figure}
\includegraphics[width=0.46\textwidth]{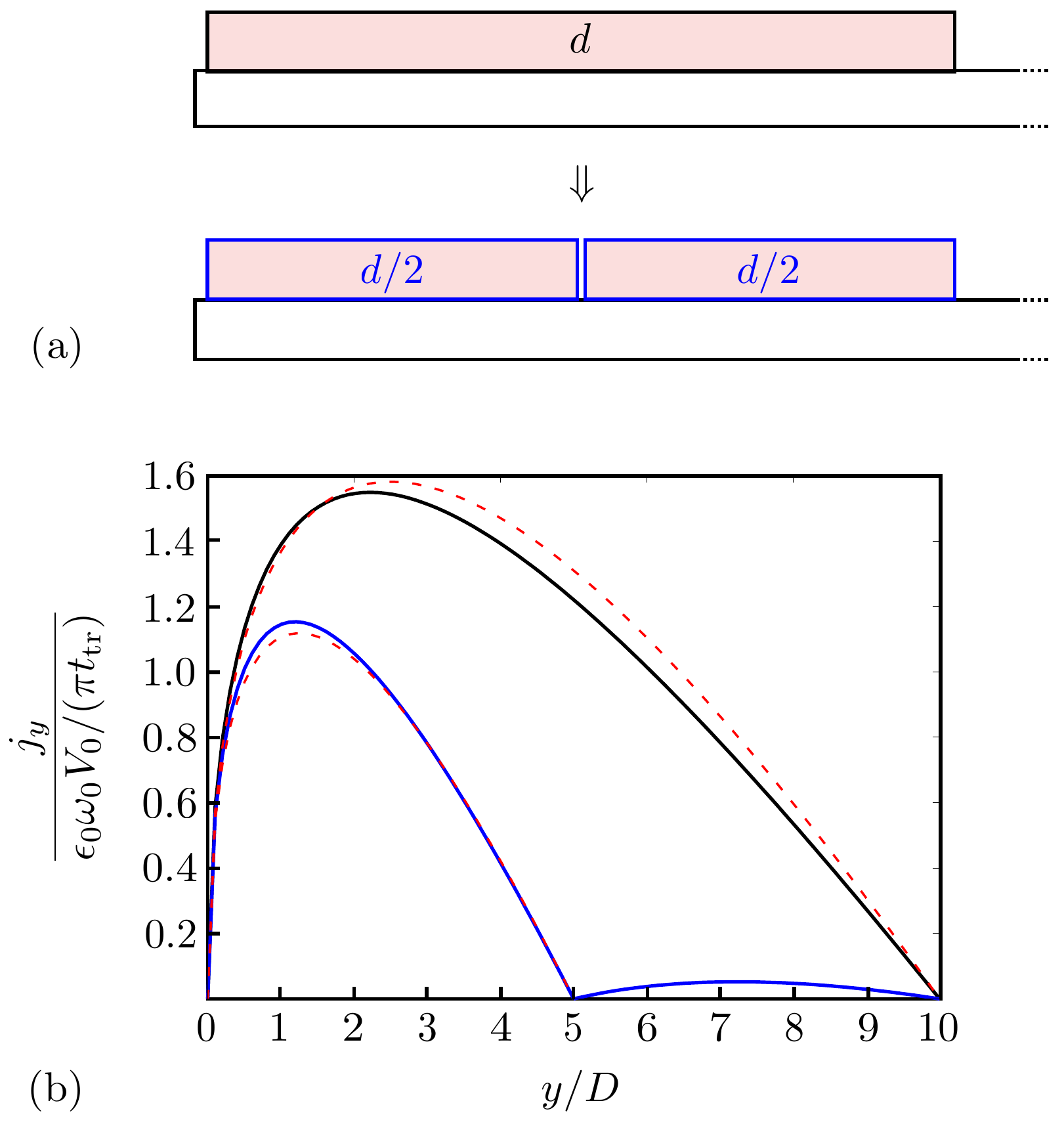}
\caption{(Color online) (a) Depiction of a trap splitting for a trap close to the capacitor edge $l=0$. The trap of size $d$ (upper graph) is divided into two smaller ones of equal size $d/2$ without direct electric contact (lower graph). (b) The dissipative current in $y$-direction inside the trap, $j_y$, as a function of $y$ for either one trap (black solid) or two split traps (size $d/2$, blue solid), for $d=10D$ and $l=0$. The current for the split trap is massively reduced in the right trap part due to the current being blocked at $y=d/2$. The dashed line corresponds to a simple approximation $j_y\sim \sqrt{y/D}(1-\sqrt{y/d})$, which approaches the exact analytic result [see Eq.~\eqref{eq:jy} in Appendix~\ref{app:Ncurr}] for $d\ll D$, but reproduces the right qualitative behavior also for $d>D$.}\label{fig:split_trap}
\end{figure}

We first focus on the worst case scenario, where a single large trap is placed at the capacitor edge close to the junction. Having a trap placed in this position may be useful in order to reduce the quasiparticle density at the junction~\cite{trapopt}. In this configuration, the losses due to currents in y-direction are dominant, see Eqs.~\eqref{eq:Py_wcs} and \eqref{eq:pxy_wsc}, and they can be efficiently reduced by splitting the original trap of size $d$ into two smaller ones each with size $d/2$, see Fig.~\ref{fig:split_trap}a. The fact that the traps are not in direct electric contact restricts the current in the $y$-direction, and in particular reduces it significantly for the second trap, which is further away from the junction, see Fig.~\ref{fig:split_trap}b. As a consequence, the dissipated power $P_y$ is approximately reduced by a factor of $2$, see Eq.~\eqref{eq:Py_wcs} (up to negligible logarithmic factors).

Therefore, trap splitting is effective not only with respect to the quasiparticle evacuation but also with respect to loss reduction. We note however a subtle difference. For quasiparticle evacuation, the trap splitting is effective only if the split traps are placed apart in order to reduce the overall diffusion time. For the losses, the only important aspect is the restriction of the harmful currents in $y$-direction: whether the split traps are placed far apart or not has a negligible effect.

This is in contrast to the regime of traps far away from the junction $l\gg d,D$, as soon as $P_x$ is dominant, see Eq.~\eqref{eq:px_f}. Here, the mere splitting of the trap does not help to reduce the overall dissipation, because, as can be expected, the insulating barrier cannot efficiently reduce the current $j_x$. Indeed, the linear dependence $P_x\sim d$ suggests that the contributions of the split traps simply add up to the same value. Hence, in order to mitigate dissipative effects in this regime, the split traps have to be placed further apart in order to capitalize on the suppression of the dissipated power with distance $l$, $P_x\sim 1/l^2$.

In summary, we find that when the trap is close to the capacitor edge, the power loss is given by the parallel component $P_y$ and that, as expected, the dissipation is reduced by making the trap smaller, cf. \esref{eq:pxy_wsc} and \rref{eq:py_se}, or splitting the trap into smaller subparts. For a trap far from the edge, it depends on the trap size which of the two component is more important. In any case, comparing
\esref{eq:px_f}-\rref{eq:py_f} to \rref{eq:pxy_wsc} and \rref{eq:py_se} we find that the dissipated power is always greatly reduced by moving the trap away from the edge.

We are now ready to estimate the impact of the trap on the qubit quality factor $Q_q$, by distinguishing a ``background'' contribution $Q^{0}$ (in the absence of trap) and a trap contribution $Q_\text{N}$ from dissipation in the normal metal [cf. \eref{QP}]:
\be
Q_q^{-1} = \left[Q^{(0)}\right]^{-1} + Q_\text{N}^{-1} \equiv \frac{1}{\omega_0 T_1^{(0)}} + \frac{P_N}{\omega_0 E}\, ,
\ee
where $T_1^{(0)}$ is the relaxation time in the absence of trap, usually of the order of tens to few hundreds $\mu$s in state-of-the-art transmons~\cite{Dial}. For a typical qubit frequency $\omega_0 = 2\pi\times7\,$GHz, we then estimate $Q^{(0)}\sim 10^6$-$10^7$. To estimate a lower bound on $Q_\text{N}$, we consider the worst case of a large trap near the edge, in which $P_\text{N}$ is given by $P_y$ of \eref{eq:pxy_wsc}, with $P_0$ of \eref{eq:P0}. The voltage $V_0$ appearing there can be estimated in terms of the qubit parameters using the Josephson relation
\be\label{eq:Jr}
2eV_0 = \omega_0 \varphi
\ee
and that, in the harmonic oscillator approximation for the transmon, the expectation value of the square of the phase in the ground state is
\be\label{eq:varphi}
\langle \varphi^2 \rangle = \sqrt{\frac{2E_C}{E_\text{J}}} = \frac{\omega_0}{2E_\text{J}}
\ee
The energy stored is $E\sim \omega_0$, therefore
\be
Q_\text{N}^{-1} = \left(\frac{\epsilon_0\omega_0}{\pi e}\right)^2 \sqrt{\frac{E_C}{8E_\text{J}}} \, \rho_\text{N} \frac{Wd}{t_\text{tr}}
\ee
Using $\epsilon_0 = 8.85\times 10^{-12}\,$F/m, $E_\text{J}/E_C=50$, and $\rho_\text{N} \sim 2\times10^{-8}\,\Omega\cdot$m~\cite{rhon} for Cu films, for a large trap with $W = 250\,\mu$m, $d=100\,\mu$m, $t_\text{tr}=80\,$nm, we find $Q_\text{N} \sim 10^{8}$.

The above estimate shows that even covering a large fraction of the transmon capacitor plates with the nor\-mal-metal trap, ohmic losses in the latter would not limit the lifetime of the best transmon currently available. On the other hand, assuming other sources of relaxation can be further reduced, such a big trap would limit $T_1$ to a few millisecond. We note that the traps that have been experimentally tested~\cite{Riwar} and that are expected to give optimal performances~\cite{trapopt} (albeit for a different device geometry) have areas $Wd$ that are 1-2 orders of magnitude smaller than that used in our estimate, and that they could be placed in regions of lower electric field, away from edges. Thus we do not expect properly designed traps to pose significant limitations to qubit lifetime even for future, improved transmons. So far we have neglected the tunnel current between superconductor and normal-metal -- in the next section we show that its effect is indeed negligible.

\section{Qubit relaxation due to N-S tunnelling}
\label{sec:decoherence}

In this Section we estimate the rates for transitions between transmon levels caused by tunnelling between N and S. We again approximate the transmon as a harmonic oscillator with frequency $\omega_0$; this justifies a semiclassical approach within which we treat the ac voltage difference between N and S as a time-dependent perturbation and use Fermi golden rule to calculate the down ($\Gamma_\downarrow$) and up ($\Gamma_\uparrow$) transition rates between the two lowest oscillator levels.

We model tunnelling with the standard tunnel Hamiltonian, and as we show in Appendix~\ref{app:vns} we can express the relaxation rate $\Gamma_\downarrow$ as a product between a squared prefactor ${\cal M}^2$ and a spectral density $S$ calculated at the qubit frequency:
\be\label{eq:gd}
\Gamma_\downarrow = {\cal M}^2 S(\omega_0)\, .
\ee
The prefactor ${\cal M}$ is proportional to the amplitude of the N-S voltage difference $V_\text{NS}$; the latter, computed as the ratio between surface charge and capacitance at the N-S interface, is proportional to $V_0$ and hence to qubit parameters [cf. \esref{eq:Jr} and \rref{eq:varphi}]:
\be\label{eq:cM}
{\cal M} \equiv \frac{e V_\text{NS}}{\omega_0} = \frac{{\cal G}}{\pi \epsilon_r} \frac{t_\text{NS}}{d} \left(\frac{E_C}{8E_\text{J}}\right)^{1/4}
\ee
where $t_\text{NS}$ is the thickness of the insulating layer between N and S and $\epsilon_r$ its dielectric constant. The geometry-dependent factor ${\cal G}$ accounts for the size and position of the trap: we have
\be\label{Gln}
{\cal G} \simeq  \ln \left(\frac{d}{\max\{l,D/4\}}\right)
\ee
for a large trap near the edge, $d \gg D,\, l$;
\be
{\cal G} \simeq 2{\sqrt{\frac{d}{D}}}
\ee
for a small trap near the edge, $l \ll d \lesssim D$; and
\be\label{Gsf}
{\cal G} \simeq \frac{d}{l}
\ee
for a trap far from the edge, $d, \,D \ll l$. For later use in our estimates for the rates, we note that since for the typical insulating material in qubits, aluminum oxide, $\epsilon_r \sim 4$~\cite{pekolaj}, even for a very large trap near the edge, $d=20D$, we have ${\cal G}/\pi\epsilon_r <1$. Also, the insulator thickness $t_\text{NS} \sim 1\,$nm is much smaller than the trap size $d$, and in a transmon the ratio $E_\text{J}/E_C$ is large.

The spectral density $S$ has two terms, accounting for tunnelling from S to N and vice versa,
\bea
S(\omega) & = & S_{\text{SN}} (\omega) + S_\text{NS} (\omega) \label{eq:S} \\
S_\text{SN}(\omega) &=& \frac{g_\text{NS}}{\pi g_K} \int_\Delta d\epsilon \frac{\epsilon}{\sqrt{\epsilon^2-\Delta^2}}  \, f_\text{S}(\epsilon) \left[1-f_\text{N}(\epsilon+\omega)\right]\nonumber \\ \\
S_\text{NS}(\omega) &=& \frac{g_\text{NS}}{\pi g_K}  \int_\Delta d\epsilon \frac{\epsilon}{\sqrt{\epsilon^2-\Delta^2}} \,
f_\text{N} (\epsilon -\omega) \left[1-f_\text{S} (\epsilon) \right] \nonumber \\
\eea
where $g_\text{NS}$ is the conductance of the N-S interface and $g_K=e^2/2\pi$ the conductance quantum. At the energies of interest (of order of the gap $\Delta$), it is reasonable to assume $f_\text{S},\, f_\text{N} \ll 1$; then we can write
\be\label{eq:SSN}
S_\text{SN}(\omega) = \frac{4E_\text{J}}{\pi} \frac{g_\text{NS}}{g_T} x_\qp
\ee
where
\be\label{eq_x_qp}
x_\text{qp} = \frac{2}{\Delta}\int_\Delta d\epsilon \frac{\epsilon}{\sqrt{\epsilon^2 - \Delta^2}}\, f_\text{S} (\epsilon)
\ee
is the quasiparticle density (normalized by the Cooper pair density) and we have used the relation $E_\text{J} = g_T \Delta/8 g_K$ with $g_T$ the conductance of the qubit Josephson junction.

The spectral density $S_\text{SN}$ is proportional to $x_\qp$, irrespective of the details of $f_\text{S}$ (but we remind the requirement $f_\text{N}(\epsilon) \ll 1$ for $\epsilon>\Delta$).
In contrast, to conveniently estimate the spectral density $S_\text{NS}$ we limit our considerations to a quasiparticle distribution possessing a characteristic energy scale, $\delta E$, such that $f_\text{S}$ rapidly vanishes at energies above $\Delta + \delta E$.
Thus we can distinguish between ``cold'' ($\delta E \ll \omega_0$) and ``hot'' ($\delta E \gg \omega_0$) quasiparticles.
We take the distribution function $f_\text{N}$ at the energies of interest to be determined by the balance between elastic NS tunnelling and relaxation in N, as in the model developed in \ocite{Riwar}. Then we can consider two limiting cases: first, if we assume fast relaxation to energies below $\Delta - \omega_0$, we can obviously neglect $S_\text{NS}$ in comparison with $S_\text{SN}$. Second, if relaxation is slow and can be disregarded, elastic tunnelling implies $f_\text{N}(\epsilon) \simeq \theta(\epsilon-\Delta)f_\text{S}(\epsilon)$ above the gap, and $f_N$ negligibly small at lower energies; then, assuming that the gap is the largest energy scale ($\Delta \gg \delta E,\, \omega_0$), we have the bound,
\begin{equation}
S_\text{NS} \lesssim \frac{\sqrt{\delta E}}{\sqrt{\delta E+ \omega_0}}S_\text{SN}\ .
\end{equation}
In the hot quasiparticle regime $\delta E \gg \omega_0$, $S_\text{NS}$ can therefore contribute a term of the same order as $S_\text{SN}$, while for cold quasiparticles the latter is dominant. In summary, we find that $S_\text{NS}$ is at most comparable to, or much smaller than, $S_\text{SN}$, so that for an order-of-magnitude estimate we can keep only the latter contribution to $S$ and we arrive at
\be
\Gamma_\downarrow \approx {\cal M}^2  S_\text{SN} (\omega_0) \, .
\ee
Similar considerations apply to the calculation of the rate $\Gamma_\uparrow = {\cal M}^2 S(-\omega_0)$, and we find $\Gamma_\uparrow \approx \Gamma_\downarrow$.

It is instructive to compare the above results with the transmon decay rate $\Gamma_\qp$ due to quasiparticle tunnelling through the Josephson junction~\cite{prb1}, which can be expressed as
\be
\Gamma_\qp = {\cal M}_\qp^2 S_\qp (\omega_0) \,,
\ee
where the matrix element for quasiparticle tunnelling is ${\cal M}_\qp = \left(E_C/8E_\text{J}\right)^{1/4}$. The spectral density $S_\qp$ takes a simple form for cold quasiparticles ($\omega \gg \delta E$),
\be
S_\qp (\omega) = \frac{8E_\text{J}}{\pi} x_{\qp,\text{J}} \sqrt{\frac{2\Delta}{\omega}} \, ,
\ee
and taking into account that the quasiparticle density at the Josephson junction $x_{\qp,\text{J}}$ is generally larger than that at the trap $x_\qp$ (and in fact considerably larger for large traps, see~\cite{trapopt}), in this regime we have
\be\label{Gdqpr}
\frac{\Gamma_\downarrow}{\Gamma_\qp} < \left(\frac{{\cal G}}{\pi\epsilon_r}\right)^2 \sqrt{\frac{2\omega_0}{\Delta}}\left[\left(\frac{t_\text{NS}}{d}\right)^2\frac{g_\text{NS}}{g_T}\right]\,.
\ee
The first two terms on the right hand side are at most of order unity [cf. discussion after \eref{Gsf}]. For the term in square bracket, we note that since the same oxide forms both the N-S and junction barrier, the ratio of conductances can be estimated as the ratio of areas, $g_\text{NS}/g_T \sim Wd/s_\text{J}^2$, where the typical junction lateral size is $s_\text{J} \sim 0.2\,\mu$m; then the factors in square brackets are $t_\text{NS}^2 W/s_\text{J}^2 d$. While the trap aspect ratio $W/d$ could be large, it realistically does not exceed $\sim 10^2$, as the width $W$ of capacitor plates is at most in the few hundred $\mu$m range and trap size $d$ is at least a few $\mu$m (note that for larger traps the aspect ratio is smaller); on the other hand, we estimate $\left(t_\text{NS}/s_\text{J}\right)^2 \sim 10^{-4}$. Collecting all factors, we conclude that $\Gamma_\downarrow/\Gamma_\qp \ll 10^{-2}$, and the N-S tunnelling contribution to the qubit $T_1$ time can be neglected.

While the bound on $\Gamma_\downarrow/\Gamma_\qp$ in \eref{Gdqpr} is restricted to the cold quasiparticle regime, we can more generally put a bound on the rate $\Gamma_\downarrow$ itself by writing it explicitly in the form
\be
\Gamma_\downarrow = \frac{\omega_0}{2\pi} x_\qp \left(\frac{{\cal G}}{\pi\epsilon_r}\right)^2 \left[\left(\frac{t_\text{NS}}{s_\text{J}}\right)^2\frac{W}{d}\right] \ll 10^{-2} \frac{\omega_0}{2\pi} x_\qp \, ,
\ee
where the bound follows from the estimates in the preceding paragraph. We stress that this bound cannot be saturated in practice: the factor ${\cal G}/\pi\epsilon_r$ is of order unity for a large trap near the capacitor edge, for which however the aspect ratio is also of order unity rather than $\sim 10^2$; on the other hand, for a small trap with large aspect ratio, ${\cal G}$ is necessarily small. Using $\Gamma_\uparrow \approx \Gamma_\downarrow$, the expression above translates into a bound on the quality factor,
\be\label{eq_q_factor_lifetime}
Q_\downarrow = \frac{\omega_0}{\Gamma_\downarrow+\Gamma_\uparrow} \gg \frac{10^2}{x_\qp}
\ee
Since even in the absence of traps the low-temperature quasiparticle density is small, $x_\qp < 10^{-5}$~\cite{wang}, and the density near the trap is expected to be much smaller, $x_\qp\sim 10^{-8}$~\cite{trapopt}, we conclude that N-S tunnelling does not significantly affect the quality factor.

In closing this section, we remark that the semiclassical approach employed here can be validated by a fully quantum mechanical calculation within the circuit model, see Appendix~\ref{app:ns_le}. Also, the approximate equality between the up and down transition rates, $\Gamma_\uparrow \approx \Gamma_\downarrow$, is a clear indication of non-equilibrium; indeed in thermal equilibrium we would have the detailed balance relation $\Gamma_\uparrow = e^{-\omega_0/T} \Gamma_\downarrow$, and therefore for ``cold'' quasiparticles $\Gamma_\uparrow \ll \Gamma_\downarrow$. In Appendix~\ref{app:ns_diss} we comment on the relationship between transition rates, junction impedance (and hence dissipated power), and thermal equilibrium.

\section{Summary}
\label{sec:summary}

In this work we study relaxation in superconducting qubits caused by normal-metal traps. By analyzing the spatially inhomogeneous electric field in typical transmon qubits (see Fig.~\ref{fig:transmon_capacitor}), we show that the dissipation due to ohmic losses inside the normal metal can be dominated by lateral currents (perpendicular with respect to the electric field), see discussion after Eq.~\eqref{eq_P_i}. As a consequence, a simple circuit picture with lumped elements, see Fig.~\ref{fig:circuit-scheme}, can fail to estimate the ohmic contribution to the quality factor. Our study indicates how to limit ohmic dissipation by appropriately choosing trap size and position, as well as by splitting the trap (Fig.~\ref{fig:split_trap}).
The second important contribution to qubit relaxation comes from tunnelling currents across the N-S interface, which we treat by means of Fermi's golden rule. To accurately estimate the quality factor, it is important to take into account the nonequilibrium distribution of quasiparticles in both superconductor and normal metal. We thus find that the excitation rate is comparable in magnitude to the relaxation rate,  a result that clearly deviates from the expectation of detailed balance in thermal equilibrium, see the last paragraph of Sec.~\ref{sec:decoherence}. 
We conclude that neither of the two contributions to the qubit relaxation provides a serious limitation to the quality factor of the best qubits available.

\acknowledgments

We gratefully acknowledge fruitful discussions with L. D. Burkhart, Y. Y. Gao, A. Hosseinhkani, and especially R. J. Schoelkopf.
This work was supported in part by the EU under REA
Grant Agreement No. CIG-618258 (G.C.), ARO Grant
W911NF-14-1-0011 and a Max Planck award (R.P.R.), and DOE contract DEFG02-08ER46482 (L.G.).

\appendix

\section{Surface charge and current densities}
\label{app:conformal}

In this Appendix, we present details of the calculation of the quantities needed to estimate ohmic losses in the trap, namely surface charge and current densities; we also take into account the finite thickness $t_\text{S}$ of the superconducting film.
We remind that the transmon is treated as
two coplanar capacitor plates at a voltage difference $V$. As explained in Sec.~\ref{sec:normal_currents}, we  assume translational
invariance in $z$-direction (plate width $W\rightarrow\infty$).
The two superconducting electrodes are then thin, semi-infinite plates of thickness
$t_\text{S}$ in the $x$-direction at a distance $D$ from each other
in $y$-direction, see Figs.~\ref{fig:transmon_capacitor} and \ref{fig:transmons}. This problem can be mapped onto a parallel plate
capacitor by using a conformal map in $x,y$-space (while leaving
$z$ invariant), as we show below.

\subsection{Infinitely thin plates}
\label{app:2Dplates}

Let us start with the limiting case $t_\text{S}\rightarrow0$. We denote the target space
of the actual physical problem (coplanar plates) with the unitless
complex variable $\tau=t/D$, where $t=y+ix$. The initial space of
the parallel plates we denote with the complex variable $\zeta$.
We assume that the lower and upper parallel plates are situated at $\text{Im}\,\zeta=0,\,1$,
respectively, while the coplanar capacitor plates are situated at $\text{Im}\,\tau=0$,
and the left (right) plate is at $\text{Re}\,\tau<-1$ ($\text{Re}\,\tau>0$). The
map that transforms between the parallel plate space $\zeta$ and the coplanar plate target space, $\tau$,
is
\be
\zeta=\frac{1}{\pi}\text{arccosh}\left(2\tau+1\right)\,.
\ee
For a voltage difference $V$ between the plates,
the potential $\widetilde{\varphi}$ in the initial space is
\be\label{eq:tvp}
\widetilde{\varphi}\left(\zeta\right)=-V\,\text{Im}\,\zeta
\ee
(the sign is chosen so that the right coplanar plate is at higher potential). Thus, in the coplanar case we find the potential
\be
\varphi\left(\tau\right)=-\frac{V}{\pi}\text{Im}\left[\text{arccosh}\left(2\tau+1\right)\right].
\ee
The surface charge density $\sigma$ on the coplanar plates is related to
the normal component of the electric field $\vec{E}=-\nabla\varphi$
at the plates,
\be
\sigma=\epsilon_{0}E_{n}.
\ee
On the right plate surface ($x \to 0^{+}$ and $y>0$) we find \eref{sigma_y}, which
for positions close to the edge, $y \ll D$, has a square root divergence,
\be\label{eq:sydiv}
\sigma\left(y\right)\approx\frac{\epsilon_{0}V}{\pi\sqrt{D}}\frac{1}{\sqrt{y}} \, ,
\ee
in agreement with \ocite{jackson}.

\subsection{Currents in the normal metal}
\label{app:Ncurr}

Having found the surface charge density, we can now calculate the components of the current density $\vec{j}$ in the normal metal, which are needed to find the dissipated power in \eref{eq_P_i}.
The current density $\vec{j}$ inside the normal metal is obtained by demanding current conservation in the bulk,
$\partial_{x}j_{x}+\partial_{y}j_{y}=0$, with the boundary conditions
\bea
j_{x}\left(t_{\text{tr}},y\right)&=&\dot{\sigma}\left(y\right)\, , \\
j_{x}\left(0,y\right)&=&-\dot{\sigma}'  ,\\
j_y(x,l)&=&j_y(x,l+d)=0 \, .
\eea
at the edges of the trap. We thus find
\begin{eqnarray}
j_{x}\left(x,y\right) & = & \frac{\epsilon_{0}\dot{V}}{\pi t_{\text{tr}}}\left[x\partial_{y}q\left(y\right)+\frac{t_{\text{tr}}-x}{d}q\left(l+d\right)\right] \label{eq:jx}\\
j_{y}\left(x,y\right) & = & \frac{\epsilon_{0}\dot{V}}{\pi t_{\text{tr}}}\left[-q\left(y\right)+\frac{y-l}{d}q\left(l+d\right)\right],
\label{eq:jy}
\end{eqnarray}
where $q\left(y\right)$ is the dimensionless integral
of the upper surface charge density $\sigma(y)$ [\eref{sigma_y}],
\be\begin{split}
q\left(y\right) & = \int_{l}^{y}dy'\frac{1}{\sqrt{y'\left(y'+D\right)}} \\ & =
\text{arccosh} \left(\frac{2y}{D}+1\right)- \text{arccosh}\left(\frac{2l}{D}+1\right).
\end{split}\ee
Equations \rref{eq:jx} and \rref{eq:jy} show that, due to the inhomogeneous field, charges within the normal metal are displaced not only in the vertical direction ($x$) but also horizontally ($y$),
see Fig.~\ref{fig:transmon_capacitor}b. Note that while the vertical current $j_x$ inherits from the surface charge $\sigma(y)$ the divergence in \eref{eq:sydiv} when $y\rightarrow l\rightarrow0$ [cf. the first term in square brackets in \eref{eq:jx}], the horizontal current is always finite.

\subsection{Finite plate thickness}
\label{app:regular}

The divergence of $\sigma$ as $y\to 0$, \eref{eq:sydiv}, is integrable.
However, the contribution $P_x$ to the dissipated power defined in \eref{eq_P_i} is obtained by integrating $j_x^2 \propto \sigma^2$, but
the integral at the edge is not well-defined.
To regularize this integral, we keep the thickness $t_\text{S}$ finite but small, $t_\text{S}\ll D$.

In principle we can account for the finite thickness via a more cumbersome conformal mapping.
To simplify the calculation for the finite
thickness problem, we consider the region of the right plate close to the edge, $y\ll D$,
which can be approximated as a single-plate capacitor at infinite
distance from ground and voltage  $\widetilde{V}$. We then establish
the relation between $\widetilde{V}$ and $V$ by appropriately matching the solution for the single-plate, finite thickness case to that of the two-plate, zero thickness one in the region
$t_\text{S}\ll y\ll D$.

For the single-plate case, the potential in the initial space $\zeta$ is as in \eref{eq:tvp},
\be\label{eq:tvp2}
\widetilde{\varphi}= -\widetilde{V}\,\text{Im}\,\zeta
\ee
However, the conformal mapping to the target space $\widetilde{\tau}=t/t_\text{S}$ is now
\be\label{eq_tau_of_zeta}
\widetilde{\tau}=\frac{1}{2\pi}\left[\sinh\left(2\text{arccosh}\zeta\right)-2\text{arccosh}\zeta\right].
\ee
with the upper and lower surfaces of the finite thickness plate
at $\text{Im}\,\widetilde{\tau}=0$ and $\text{Im}\,\widetilde{\tau}=-i$, respectively, while
the edge is at $\text{Re}\,\widetilde{\tau}=0$.

Let us consider the limit of large distances away from the edge, $y\gg t_\text{S}$.
For large absolute values of $\zeta$, we may approximate
\be
\text{arccosh}\zeta\approx\ln\left(2\zeta\right).
\ee
Using this approximation in \eref{eq_tau_of_zeta} together with $\sinh x \approx e^x/2$ for large $x$, and solving for $\zeta$, \eref{eq:tvp2} gives us
\be
\varphi\approx-\sqrt{\pi}\widetilde{V}\text{Im}\sqrt{\frac{t}{t_\text{S}}}
\ee
for the potential in the target space. We can match this approximate solution to that in \eref{eq:tvp} for $\tau \ll 1$
\be
\frac{2V}{\pi}\text{Im}\sqrt{\frac{t}{D}}=\sqrt{\pi}\widetilde{V}\text{Im}\sqrt{\frac{t}{t_\text{S}}}
\ee
by setting
\be
\widetilde{V}=\frac{2V}{\pi}\sqrt{\frac{t_\text{S}}{\pi D}}.
\ee


Having determined the prefactor $\widetilde{V}$, let us now consider in more detail the potential near the edge as determined by \esref{eq:tvp2} and \rref{eq_tau_of_zeta}. For $|\tilde{\tau}|\ll 1$, we can approximately invert the latter,
\be
\zeta\approx 1+\frac{1}{2}\left(\frac{3\pi}{2}\frac{t}{t_\text{S}}\right)^{2/3}
\ee
and substituting into the former we obtain
\be
\varphi\left(t\right)\approx-\frac{V}{\pi^{3/2}}\sqrt{\frac{t_\text{S}}{D}}\left(\frac{3\pi}{2}\right)^{2/3}\text{Im}\left[\left(\frac{t}{t_\text{S}}\right)^{2/3}\right].
\ee
This results in a surface charge density on the top of the plate,
\be
\sigma\left(y\right)=\frac{\epsilon_{0}V}{\sqrt{\pi t_\text{S} D}}\left(\frac{2}{3\pi}\frac{t_\text{S}}{y}\right)^{1/3}.
\ee
that diverges as $y^{-1/3}$ (cf. \ocite{jackson}). This weaker divergence than that in \eref{eq:sydiv} demonstrates that
the integral of $\sigma^{2}$ is in fact finite.

Indeed, let us consider the integral
\be
\int_{0}^{b}dy \, \sigma^{2}\left(y\right) = \int_{0}^{a}dy\,\sigma^{2}\left(y\right)+\int_{a}^{b}dy\,\sigma^{2}\left(y\right)
\ee
with $a, \, b\gg t_\text{S}$ and $a\ll D$. For the second integral on the right hand side, we can use \eref{sigma_y} to find
\be\label{eq:s2ab}
\int_{a}^{b}dy \, \sigma^{2}\left(y\right)\approx\frac{\epsilon_{0}^{2}V^{2}}{\pi^{2}D}\ln\left[\frac{bD}{\left(b+D\right)a}\right].
\ee
This expression would of course diverge logarithmically for $a\to 0$.
For the integral between 0 and $a$, the appropriate approximate expression for the surface charge density derived from \esref{eq:tvp2} and \rref{eq_tau_of_zeta} is
\begin{eqnarray}
\sigma\left(y\right) & = & -\epsilon_{0}\partial_{x}\varphi\left(y\right)\\
 & = & \frac{\epsilon_{0}V}{\sqrt{\pi Dt_\text{S}}}\frac{1}{\sqrt{\zeta^{2}\left(y\right)-1}}
\end{eqnarray}
and using $a\gg t_\text{S}$ we arrive at
\be\label{eq:s20a}
\int_{0}^{a}dy\,\sigma^{2}\left(y\right)
\approx \frac{\epsilon_{0}^{2}V^{2}}{\pi^{2}D}\ln\left(\frac{4\pi a}{t_\text{S}}\right).
\ee
Summing together \esref{eq:s2ab} and \rref{eq:s20a}, we obtain
\be
\int_{0}^{b}dy\,\sigma^{2}\left(y\right)=\frac{\epsilon_{0}^{2}V^{2}}{\pi^{2}D}\ln\left[\frac{bD}{\left(b+D\right)t_\text{S}/4\pi}\right],
\ee
and thus we see that the logarithmic divergence is cut off by $t_\text{S}/4\pi$.

\section{Lumped-element approach for small, far traps}
\label{app:lest}

We consider here the question of when a simple lumped element description is appropriate. One may expect it to be valid when the trap size is the smallest length scale in the problem; this happens in the regime of small trap far from the edge, $d, D\ll l$. In a lumped element description, we write the dissipated power as
\be\label{PNle}
P_\text{N} = \frac12 R_\text{N} I_\text{N}^2
\ee
with $I_\text{N}$ denoting the current flowing through the normal metal. The assumption of small trap size means we take the current to move only perpendicularly to the N-S interface, neglecting variation of the charge density in the $y$ direction; from the considerations after \eref{eq:py_f} in Sec.~\ref{sec:normal_currents}, we know this is correct for $d \ll \sqrt{11 t_\text{tr} l}$. The current is given by the time derivative of the surface charge density times area of the trap,
\be
I_\text{N} = \dot\sigma' W d \, ,
\ee
and for the resistance we have
\be
R_\text{N} = \rho_\text{N} t_\text{tr}/W d \, .
\ee
Using \esref{sigma_y} and \rref{eq:sigmap} we find [see also \esref{sigmapG} and \rref{Gsf}]
\be
\dot\sigma' = -\frac{\epsilon_0 \omega_0 V_0}{\pi d} \frac{d}{l}\,,
\ee
and substituting the last three equations into \eref{PNle} we indeed recover \eref{eq:px_f}.
We stress that for the other regimes considered (small and large trap near the edge), the simple description fails.

For later use in Appendix~\ref{app:ns_le}, we also calculate here the value of capacitance $C_\text{N}$ between trap and the transmon capacitor plate to which it is not in tunnel contact, cf. Fig.~\ref{fig:circuit-scheme}. Adopting again a simple approach, the capacitance is calculated as the ratio between charge and voltage,
\be\label{eq:CNfs}
C_\text{N} = \frac{|\sigma'| Wd}{V_0} = \frac{\epsilon_0 Wd}{\pi l} \, .
\ee
This expression correspond to the capacitance of a parallel plate capacitor of area $Wd$ with plates at distance $\pi l$, the factor $\pi$ accounting here for an effective distance given by half a circle of radius $l\gg D$ (cf. the field lines in Fig.~\ref{fig:transmon_capacitor}a).

\section{Qubit relaxation due to $V_\text{NS}$}
\label{app:vns}

This Appendix outlines the derivation of \eref{eq:gd} within a semiclassical approach.
Our starting point is the following Hamiltonian in the excitation representation
\bea
H &=& H_{eh}+H_\qp+H_T \label{eq:H} \\
H_{eh} &=& \sum_{n,\sigma} \left|\xi_n\right| c^\dagger_{n\sigma}c_{n\sigma} \label{Heh}
\\
H_\qp &=& \sum_{m,\sigma} \epsilon_m \alpha^\dagger_{m\sigma} \alpha_{m\sigma}
\\
H_T &=& \tilde{t} \sum_{n,m,\sigma} \left(e^{i\phi \,\text{sgn}(n)}\tilde{u}_{mn} c^\dagger_{n\sigma}\alpha_{m\sigma} +\text{H.c.} \right)\label{HT}
\eea
where $c_{n\sigma}$ are annihilation operators for electron-like (hole-like) excitations in the normal metal above (below) the Fermi level, $n>0$ ($n<0$), $\alpha_{m\sigma}$ are annihilation operators for quasiparticle excitations in the superconductor, $\xi_n$ are the single-particle energy levels in the normal metal, $\epsilon_m=\sqrt{\xi_m^2+\Delta^2}$ with $\xi_m$ the single-particle energy levels in the superconductor of gap $\Delta$, and $\tilde{t}$ is the tunnelling amplitude.  We also define the Bogoliubov amplitudes
\be
\tilde{u}_{nm} = \text{sgn}(n) \, \frac{1}{\sqrt{2}}\sqrt{1+\text{sgn}(n) \frac{\xi_m}{\epsilon_m}}.
\ee
These expressions can be obtained by taking the limit of zero gap ($\Delta\to0$) in one of the two superconductors forming an S-I-S junction, as considered \textit{e.g.} in \ocite{prb1}.

The time-dependent part of the phase difference $\phi$ in \eref{HT} is related to the ac voltage,
\be
\phi = \frac{e V_\text{NS}}{\omega_0} \sin (\omega_0 t) \, .
\ee
where $V_\text{NS}$ is the amplitude of the voltage difference.
We expand the exponential up to first order in $e V_\text{NS}/\omega_0$ and treat the first order term as the perturbation causing transitions. Fermi golden rule then gives
\be\label{Gammadown}\begin{split}
\Gamma_\downarrow = &\,2\pi \tilde{t}^2 \left(\frac{eV_\text{NS}}{\omega_0}\right)^2 2 \sum_{n,m} \tilde{u}_{nm}^2 \\ &\times
\Big[f_\text{S}(\xi_m) \left(1-f_\text{N}(\xi_n)\right) \delta\left(\epsilon_m +\omega_0 - |\xi_n|\right) \\ &+
f_\text{N}(\xi_n) \left(1-f_\text{S}(\xi_m)\right)\delta\left(|\xi_n| +\omega_0 - \epsilon_m\right)\!\Big]
\end{split}\ee
where the factor 2 in front of the sum originates from the sum over spin direction $\sigma$ and the distribution functions appear because we average over the initial state of the excitations in the superconductor (for the first term in square bracket) or in the normal metal (for the second term). Introducing as usual the density of states to transform the sums into integrals, and eliminating one of the integrals thanks to the delta-functions, we arrive at
\be\begin{split}
& \Gamma_\downarrow = \frac{g_\text{NS}}{\pi g_K} \left(\frac{eV_\text{NS}}{\omega_0}\right)^2 \int_\Delta d\epsilon \frac{\epsilon}{\sqrt{\epsilon^2-\Delta^2}} \\ & \quad \left[f_\text{S}(\epsilon) \left(1-f_\text{N}(\epsilon+\omega_0)\right) +
f_\text{N} (\epsilon -\omega_0) \left(1-f_\text{S} (\epsilon) \right)\right]
\end{split}\ee
where $g_\text{NS} = 4\pi e^2 \tilde{t}^2 \nu_\text{N} \nu_\text{S}$ is the tunnelling conductance between N and S and $g_K=e^2/2\pi$ is the conductance quantum. (Here we have assumed no charge imbalance.) This formula gives \eref{eq:gd} with the definitions in \esref{eq:cM} and \rref{eq:S}. The up rate $\Gamma_\uparrow$ is obtained with the replacement $\omega_0 \to -\omega_0$.

Next, we want to relate the voltage difference $V_\text{NS}$ to the device properties, as in the last term in \eref{eq:cM}. To this end, we remind that the surfaces at the N-S interface form a parallel plate capacitor, so we can relate voltage to charge and capacitance:
\be\label{VNS}
V_\text{NS} = \sigma' Wd/C_\text{NS}
\ee
with $\sigma'$ of \eref{eq:sigmap} and
\be\label{CNS}
C_\text{NS} = \epsilon \frac{Wd}{t_\text{NS}} \, ,
\ee
where $\epsilon$ is the permittivity of the insulating layer of thickness $t_\text{NS}$ separating the trap from the superconductor. We write the charge density $\sigma'$ as
\be\label{sigmapG}
\sigma' = \sigma_0 {\cal G} \, , \quad \sigma_0 = \frac{\epsilon_0 V_0}{\pi d}
\ee
where the factor ${\cal G}$ is a dimensionless function that accounts for the device geometry and takes the approximate forms given in \esref{Gln}-\rref{Gsf} for different regimes of trap size and position.
Substituting \esref{CNS} and \rref{sigmapG} into \eref{VNS} we find
\be\label{VNSf}
V_\text{NS} = \frac{{\cal G}}{\pi \epsilon_r} \frac{t_\text{NS}}{d} V_0
\ee
with $\epsilon_r=\epsilon/\epsilon_0$ the dielectric constant. Using \esref{eq:Jr} and \rref{eq:varphi} to express $V_0$ in terms of qubit parameters, we arrive at the last equality in \eref{eq:cM}.

\section{Qubit relaxation within the lumped element model}
\label{app:ns_le}

In this Appendix we reconsider the effect of tunnelling through the N-S interface on qubit coherence within the lumped element model introduced in Sec.~\ref{sec:circuit_pic}.
Starting from the Lagrangian ${\cal L}$ of \eref{eq:lagr}, it is straightforward to obtain the circuit Hamiltonian $H_0$. For reasons that will become clear shortly, before performing the Legendre transform, we perform a rescaling, $\phi \to 2\phi$; then in the regime $C_\text{NS} \gg C_\text{N}$, $H_0$ takes the form
\be
H_0 = 4 E_C n^2 -E_\text{J} \cos\varphi + \tilde{E}_C q^2 + 4 E_C \frac{C_\text{N}}{C_\text{NS}} n q
\ee
with $E_C = e^2/2(C+C_\text{N})$ and $\tilde{E}_C = e^2/2C_\text{NS}$. Here $n$ and $q$ are conjugate variables to $\varphi$ and the rescaled $\phi$, respectively.

We can rewrite this Hamiltonian as
\be
H_0 = 4E_C \left(n+\frac{C_\text{N}}{C_\text{NS}}\frac{q}{2}\right)^2 - E_\text{J} \cos\varphi + \bar{E}_C q^2
\ee
with $\bar{E}_C = \tilde{E}_C \left(1- C_\text{N}^2/C_\text{NS}(C+C_\text{N})\right) \approx \tilde{E}_C$. In this form it becomes evident that the charge on the normal island acts on the transmon as an offset charge $n_g$, see \ocite{koch2007}, with
\be\label{eq:ng}
n_g = q\,C_\text{N}/2C_\text{NS}.
\ee
Therefore we denote the eigenstates of this Hamiltonian by $|q,i\rangle$, where $q=0, \, \pm1,\, \pm2,\ldots$ is (excess) charge on the normal metal island and $i=0,\,1,\,2,\ldots$ the transmon state for the given charge $q$. In other words, in terms of variables $\varphi$ and $\phi$, we have
\be\label{states}
|q,i\rangle = e^{iq\phi}\Psi_{i|q}\left(\varphi\right)
\ee
where $\Psi_{i|q}$ can be written exactly in terms of Mathieu functions~\cite{koch2007}. For calculations, however, it is more practical to use the approximate tight binding wavefunctions constructed in Appendix~B of \ocite{prb1}, which in the present notation read
\be
\Psi_{i|q}\left(\varphi\right) = e^{i n_g \varphi}\frac{1}{\sqrt{N}}\sum_k \psi_i(\varphi-2\pi k) e^{-in_g2\pi k},
\ee
with $n_g$ related to $q$ by \eref{eq:ng}, $\psi_i(\varphi)$ the $i$-th wavefunction in a single well of the transmon periodic potential, and $N\gg 1$ the number of wells considered. (Here we neglect for simplicity the odd states, as N-S tunnelling can be shown to induce transitions only between sate with the same parity.)

To incorporate the effect of tunnelling at the N-S interface, we consider the total Hamiltonian $H_{tot}$ given by
\be
H_{tot} = H_0 + H
\ee
with $H$ of \eref{eq:H} (the quantity $\phi$ in \eref{HT} should be understood as an operator).

The calculation of qubit transition rates then proceed as in \ocite{prb1} and we find
\be\label{G10}
\Gamma_{10} = {\cal M}_\text{NS}^2
S \left(\omega_{0}\right)
\ee
for the relaxation rate and
\be
\Gamma_{01} =  {\cal M}_\text{NS}^2
S \left(-\omega_{0}\right)
\ee
for the excitation rate. In these formulas the use of the spectral function of \eref{eq:S} is justified if the charging energy $\tilde{E}_C$ of the normal island can be
neglected. For a small trap of area $Wd \sim (10\mu\text{m})^2$, using \eref{CNS} with $\epsilon \sim 4\epsilon_0$ and $t_\text{NS} \sim 1\,$nm, we estimate $C_\text{NS} \sim 3.5\,$pF and hence $\tilde{E}_C/h \sim 5\,$MHz (the charging energy of bigger traps is of course smaller than this). The energy $\tilde{E}_C$ is therefore smaller than even the typical fridge temperature and can be safely neglected.

For the matrix element ${\cal M}_\text{NS}$ we have
\be\label{eq:MNS}
{\cal M}_\text{NS}^2 = \left|\left\langle 0 \left| \sin\left(\frac{C_\text{N}}{C_\text{NS}} \frac{\varphi}{2} \right) \right|1 \right\rangle\right|^2 \simeq
\left(\frac{C_\text{N}}{C_\text{NS}}\right)^2 \sqrt{\frac{E_C}{8E_\text{J}}} \, ,
\ee
where here state $|i\rangle$  denotes  the state $\psi_i(\varphi)$ in a single potential well of the transmon. A number of steps are necessary to arrive at this expression.
Indeed, it is clear from \eref{HT} that the matrix element entering Fermi golden rule is in fact of the form
\be
\langle r, j | e^{\pm i\phi} | q, i \rangle
\ee
with the states of \eref{states}. For the matrix element not to vanish, we need $r=q\pm1$; this implies a term proportional to $\tilde{E}_C$ in the energy difference between initial and final states, but we neglect this contribution as explained above. After integrating over variable $\phi$ the above matrix element becomes explicitly
\be\begin{split}
\int\!d\varphi \, e^{\mp i\frac{C_\text{N}}{C_\text{NS}}\frac{\varphi}{2}} \frac{1}{N} \sum_{k,l} \psi_i (\varphi-2\pi k) \psi_j (\varphi-2\pi l) \\
\times e^{-iq \frac{C_\text{N}}{C_\text{NS}} \pi k} e^{i(q\pm1) \frac{C_\text{N}}{C_\text{NS}} \pi l},
\end{split}\ee
and keeping only the leading contribution originating from the same well, $k=l$, we find
\be
\int\!d\varphi \, e^{\mp i\frac{C_\text{N}}{C_\text{NS}}\frac{\varphi}{2}} \psi_i (\varphi) \psi_j (\varphi) \,.
\ee
For the states $i=1$ (which is antisymmetric in $\varphi$) and $j=0$ (symmetric), this expression coincides with \eref{eq:MNS}.

Let us now show that the result of this Appendix agrees, in the appropriate regime, with that in Sec.~\ref{sec:decoherence}; since the spectral densities in the expressions \rref{eq:gd} and \rref{G10} for the rates are the same, we only need to compare the matrix elements. The calculation presented here is based on a circuit model, so we expect the two approaches to be equivalent when the lumped element description is valid. As discussed in Appendix~\ref{app:lest}, this is the case for sufficiently small traps far from the capacitor edge; then the capacitance $C_\text{N}$ is given by \eref{eq:CNfs}. Using \eref{CNS} for $C_\text{NS}$, we find for their ratio
\be
\frac{C_\text{N}}{C_\text{NS}} = \frac{1}{\pi \epsilon_r} \frac{t_\text{NS}}{l}\, ,
\ee
which, using \esref{eq:cM}, \rref{Gsf}, and \rref{eq:MNS}, indeed proves the equality ${\cal M} = {\cal M}_\text{NS}$ in the small, far trap regime.

Finally, we comment on the semiclassical vs quantum circuit approach. The derivation in this Appendix on one hand justifies the semiclassical calculations of Sec.~\ref{sec:decoherence} and Appendix~\ref{app:vns}; on the other, it shows possible limitations of the lumped element, circuit description for certain devices, \textit{e.g.} with large traps. For the weakly anharmonic transmon both approaches are possible, but the semiclassical one cannot be used for highly anharmonic qubit, while the circuit one can in principle be extended to any qubit type (when use of a lumped-element model for the device is appropriate).

\section{Dissipation by current through the N-S junction}
\label{app:ns_diss}

As we point out at the end of Sec.~\ref{sec:circuit_pic}, in thermal equilibrium we expect a circuit-based calculation of the qubit quality factor to be applicable. By contrast, our estimate for the qubit transition rates due to N-S tunnelling do not conform to thermal equilibrium expectations, see the end of Sec.~\ref{sec:decoherence} -- this is not surprising since in that Section excitations were assumed to be out of equilibrium. In this Appendix we show that in non-equilibrium states we indeed cannot estimate the qubit quality factor [\eref{Qq}] using the circuit one [\eref{QP}] and that, on the contrary, this would appropriate in equilibrium.

The power $P_\text{NS}$ dissipated due to the tunnel current at the N-S junction, is given by
\be\label{eq_PNS}
P_{\text{NS}}=\frac{1}{2}\text{Re}\left[Y_{\text{NS}}\left(\omega\right)\right]V_\text{NS}^2 \, ,
\ee
where $Y_\text{NS}$ is the admittance of the N-S junction, and $V_\text{NS}$ is the amplitude of the voltage drop across the insulating barrier [\eref{VNSf}].
In the circuit picture of Sec.~\ref{sec:circuit_pic}, we identify $\text{Re}\left[Y_{\text{NS}}\left(\omega_0\right)\right]$ with $R_\text{NS}^{-1}$.
For a tunnel N-S junction in the linear response regime, the real part of the admittance can be easily computed by noting that the power is energy exchanged (positive or negative) times the rate of exchange~\cite{prb1}; in our notation this is
\be
P_\text{NS} = \omega \left(\Gamma_\downarrow - \Gamma_\uparrow \right) \, .
\ee
Using \esref{eq:gd}, \rref{eq:cM}, and \rref{eq:S}, we arrive at
\be\label{eq_Re_Y_NS}
\begin{split}
\text{Re}\left[Y_\text{NS}\left(\omega\right)\right]=\frac{4g_{NS}}{\omega}\int_{\Delta}^{\infty}\!d\epsilon\,\frac{\epsilon}{\sqrt{\epsilon^{2}-\Delta^{2}}}\\
\times\left[f_\text{N}\left(\epsilon-\omega\right)-f_\text{N}\left(\epsilon+\omega\right)\right].
\end{split}
\ee
The distribution function in the normal metal $f_\text{N}$ is shifted by either $\pm\omega$
to account for energy absorption or emission -- this structure is similar to that for the dc response, where the dc bias takes the place of the frequency $\omega$, see \ocite{Tinkham}. It is shown there that
in the absence of charge imbalance in the superconductor, there are
no direct contributions due to the distribution function $f_\text{S}$ in the superconductor.

As an example of out-of-equilibrium situation, let us consider the case of fast relaxation in the normal metal, as described in Sec.~\ref{sec:decoherence}; in this case we have $f_\text{N}(\epsilon) = \theta(\epsilon - \Delta)f_\text{S}(\epsilon)$. We further assume ``cold'' quasiparticles, with characteristic energy $\delta E$ above the gap small compared to $\omega$, so that in \eref{eq_Re_Y_NS} we can neglect the contribution of $f_\text{N}(\epsilon + \omega)$, and bound the one from $f_\text{N}(\epsilon-\omega)$ to get
\be
\text{Re}\left[Y_\text{NS}\left(\omega\right)\right] \lesssim \frac{2g_{NS}\Delta}{\omega}x_\qp \sqrt{\frac{\delta E}{\omega}}
\ee
We can now take the ratio between the ``circuit'' inverse quality factor, $ Q^{-1} = P_\text{NS}/\omega_0 E$ (with $E\sim\omega_0$), and the qubit inverse quality factor, $Q_q^{-1}\sim \Gamma_\downarrow/\omega_0$; using for $\Gamma_\downarrow$ the results of Sec.~\ref{sec:decoherence} we find
\be
\frac{Q_q}{Q} \lesssim \sqrt{\frac{\delta E}{\omega_0}} \ll 1
\ee
This bound shows that one can grossly underestimate the impact of a decay mechanism on qubit lifetime if using the dissipated power in out of equilibrium situations.

Let us now consider the case of thermal equilibrium. By comparing \esref{eq:S} and \rref{eq_Re_Y_NS}, where the distribution functions all take the Fermi-Dirac form at a given temperature $T$, one can verify the fluctuation-dissipation relation
\be
S(\omega)+S(-\omega) = \omega \coth\left(\frac{\omega}{2T}\right) \frac{1}{4\pi g_K}\text{Re}\left[Y_\text{NS}\left(\omega\right)\right]
\ee
From this identity, it follows that in thermal equilibrium we can calculate the qubit quality factor from the circuit one, since $Q_q = Q/\coth(\omega_0/2T)$. In particular, the two quality factors coincide at low temperature $T \ll \omega_0$.

\bibliographystyle{apsrev4-1}
\bibliography{biblio_qp_trap}

\end{document}